\documentclass[cameraready]{Interspeech}

\usepackage{booktabs}
\usepackage{multirow}
\usepackage{makecell}
\usepackage{textcomp}
\usepackage{xcolor}
\usepackage{graphicx}
\usepackage{caption}
\usepackage{subcaption}
\usepackage{comment}

\title{Dynamic Prosody Prediction in LLM-based TTS for Improving \\ Speaker Similarity}

\author[affiliation={1}]{Zhenwei}{Mou}
\author[affiliation={1}]{Liping}{Chen}
\author[affiliation={2}]{Yajun}{Hu}
\author[affiliation={1}]{Zhen-Hua}{Ling}
\author[affiliation={2}]{Xin}{Fang}
\author[affiliation={2}]{Jianqing}{Gao}

\address{
    $^1$ University of Science and Technology of China, Anhui, China \\
    $^2$ iFLYTEK, Anhui, China 
}

\email{zwmu@mail.ustc.edu.cn, \{lipchen, zhling\}@ustc.edu.cn, \{yjhu,xinfang,jqgao\}@iflytek.com \thanks{\emph{Corresponding author: Liping Chen.}} \thanks{This work was supported in part by the National Key Research and Development Program Project 2024YFE0217200, the Innovation and Technology Fund of the Hong Kong SAR MHP/048/24, and the National Natural Science Foundation of China under Grant 62506349 and U23B2053.}}


\usepackage{comment}

\begin{document}

\maketitle

\begin{abstract}
    Personalized text-to-speech (TTS) aims to clone the target speaker in the synthesized speech, imitating both the voice and speaking style. Current large language model (LLM)-based TTS methods ignore the style-specific prosodic patterns in generated speech, resulting in deficient style learning and thus limiting speaker similarity in synthesized speech. To this end, we investigate the prosody learning conditioned on the synthesized speech, and propose to predict the prosody of the current syllable based on previously predicted speech. Experimental results obtained on three datasets demonstrated the efficacy of the proposed dynamic prosody prediction method in enhancing the prosody learning capability, thereby improving the speaker similarity of the generated speech. Audio samples are available at \url{https://muzw.github.io/dynapros/}.
\end{abstract}

\vspace{-0.2em}
\section{Introduction}
\vspace{-0.2em}
\label{sec:intro}

Personalized text-to-speech (TTS) aims to clone the speech of a target speaker, imitating both the voice and speaking style \cite{casanova2022yourtts,xue2021cycle,skerry2018towards}. Recent advances in large language models (LLMs) have enabled their application to personalized TTS \cite{wang2023neural,guo2024fireredtts,kharitonov2023speak,du2024cosyvoice,jiang2023mega,xin2024rall,lajszczak2024base,vevo}. In LLM-based TTS models, continuous speech signals are quantized into tokens, and the target speech tokens are generated using the generative pre-trained Transformer (GPT) mechanism \cite{brown2020language}. Besides, non-autoregressive models combined with the mask strategy have also been explored\cite{wang2024maskgct,chen2024f5}. Empowered by the generative capabilities of the architectures and the large-scale speech datasets, the similarity to the target speaker of synthesized speech has been significantly improved.

\begin{figure}[!t]
    \centering
    \begin{minipage}[b]{0.4\textwidth}
        \includegraphics[width=\textwidth]{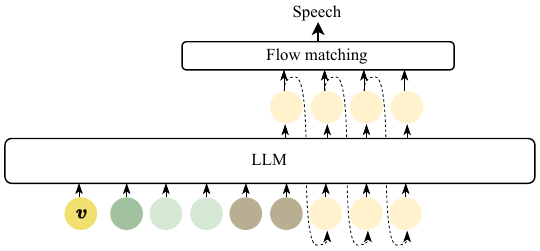}
        \vspace{-1.5em}
        \subcaption{CosyVoice LLM.}
        \label{cosyvoice}
    \end{minipage}
        
    \begin{minipage}[b]{0.4\textwidth}
        \includegraphics[width=\textwidth]{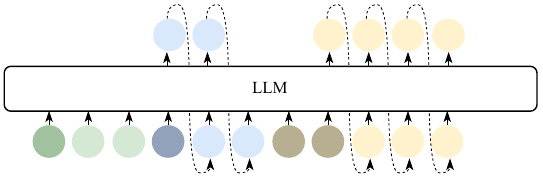}
        \vspace{-1.5em}
        \subcaption{CoT prompting.}
        \label{cot}
    \end{minipage}

    \begin{minipage}[b]{0.45\textwidth}
        \includegraphics[width=\textwidth]{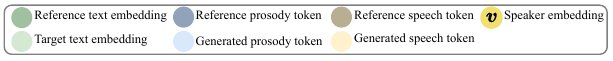}
    \end{minipage}
    \vspace{-1em}
    \caption{Illustration of the LLM in CosyVoice\cite{du2024cosyvoice} and the CoT prompting\cite{xin2024rall}.}
    \vspace{-1em}
    \label{background}
\end{figure}

\begin{figure*}[!t]
    \centering
    \vspace{-2em}
    \begin{minipage}[b]{1\textwidth}
        \centering
        \includegraphics[scale=1]{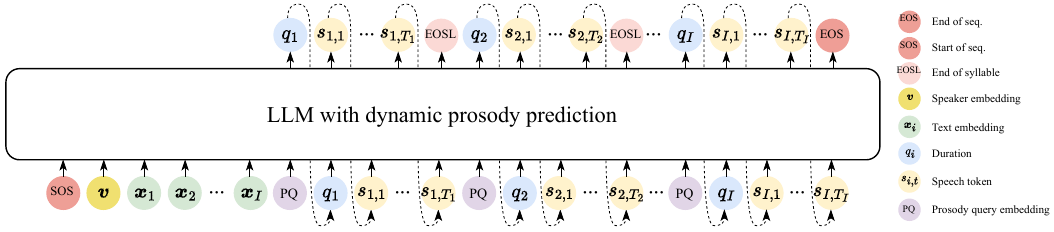} 
        \vspace{-0.8em}
        \subcaption{Overall architecture.}
        \label{fig:pe_llm}
    \end{minipage}
    
    
    \begin{minipage}[b]{0.9\textwidth}
        \centering
        \includegraphics[width=\textwidth]{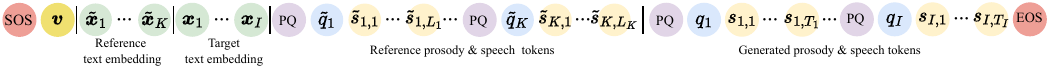}
        \vspace{-0.8em}
        \subcaption{Input and generated sequences during inference.}
        \label{fig:sequence}
    \end{minipage}
    \vspace{-0.8em}
    \caption{Illustration of the LLM architecture integrated with dynamic prosody prediction.}
    \vspace{-2em}
    \label{fig: proposed}
\end{figure*}

Existing methods\cite{wang2023neural,guo2024fireredtts,kharitonov2023speak,wang2024maskgct,chen2024f5} model speech attributes (e.g., linguistic content, voice, and speaking style) holistically from a reference utterance.
As speaker similarity is decided by both voice and speaking style, the similarity of synthesized speech to the target speaker is limited by the implicit, thus inadequate, modeling of these attributes. In CosyVoice \cite{du2024cosyvoice}, the speaker attribute is extracted from the reference speech and represented as a fixed-length embedding. Prosody, which characterizes speaking style, is explicitly predicted for target speech in \cite{jiang2023mega,xin2024rall,vevo}. In BASE TTS\cite{lajszczak2024base}, the voice and prosody attributes are modeled and represented with respective embeddings.

Among methods that explicitly model the voice or speaking style attribute, CosyVoice\cite{du2024cosyvoice} lacks the explicit modeling of speaking style. In BASE TTS\cite{lajszczak2024base}, the prosody embedding is extracted from the reference speech to provide style information for the target speech, disregarding the stylistic information conveyed by the target text. In \cite{jiang2023mega,xin2024rall,vevo}, the target prosody is predicted from both the target text and reference speech, using the chain-of-thought (CoT) prompting\cite{CoT}. Both methods statically pre-compute prosody for the entire utterance prior to synthesis, consequently neglecting style specific to target speech. To improve speaker similarity in synthesized speech through enhanced style learning, this study integrates target speech into prosody prediction by dynamically estimating syllable-level prosody conditioned on prior predicted target speech. Specifically, before generating the speech frames for a specific syllable, its prosody is predicted from the target text, reference speech, and previously generated speech, which is then used as a condition to generate the speech of the syllable. Experiments conducted on Chinese datasets within the CosyVoice framework \cite{du2024cosyvoice} demonstrated that:

1) Compared to both the baseline CosyVoice without explicit prosody modeling and the CosyVoice integrated with CoT prompting, the proposed method further enhanced the speaker similarity of the generated speech to the target speaker.

2) The proposed dynamic prosody prediction method exhibited the potential to mitigate the prosody learning gap between small- and large-scale training data.


\vspace{-0.2em}
\section{Background}
\vspace{-0.2em}
\label{sec:related}
As illustrated in Fig. \ref{background}(\subref{cosyvoice}), in the LLM of CosyVoice, a speaker embedding vector \({\boldsymbol{v}}\) is extracted from the reference speech. The input text sequence, comprising both target and reference texts, is encoded into an embedding sequence. The speech tokens are extracted from the reference speech. Using the reference speaker embedding, the text tokens and reference speech tokens as input, the LLM generates speech tokens for the target speech frames, which are then synthesized into speech by a flow matching module. In this framework, speech tokens are predicted from the target text and reference speech without explicit prosody modeling.

Explicit prosody prediction at the phoneme level using the CoT prompting was introduced in RALL-E\cite{xin2024rall} and Vevo1.5\cite{vevo}. 
As shown in Fig. \ref{background}(\subref{cot}), with the integration of the CoT strategy, all phoneme-level prosody tokens for the entire speech are predicted first, conditioned on input text and reference prosody tokens, before autoregressively generating speech tokens. This results in static pre-computation of target speech prosody prior to synthesis.

\vspace{-0.2em}
\section{Proposed method}
\vspace{-0.2em}
\label{sec:method}
This section introduces a dynamic prosody prediction strategy that incorporates previously predicted speech into prosody prediction.
\vspace{-0.2em}
\subsection{Overall Architecture}
\vspace{-0.2em}

Our study is conducted in the CosyVoice framework \cite{du2024cosyvoice} with the overall architecture illustrated in Fig. \ref{fig: proposed} (\subref{fig:pe_llm}). The input text is tokenized at the syllable level, yielding a text embedding sequence, denoted as $\boldsymbol{X}=\left\{{\boldsymbol{x}}_1,{\boldsymbol{x}}_2,...,{\boldsymbol{x}}_I\right\}$ where $I$ is the number of syllables. Assume a prosody token set ${\mathcal C}^{\rm p}$, where the superscript $^{\rm p}$ represents prosody. Utilizing the reference speaker embedding $\boldsymbol{v}$ and the text embedding sequence $\boldsymbol{X}$, the LLM predicts the probability distribution over ${\mathcal C}^{\rm p}$ for the $i$-th syllable $\boldsymbol{y}_i$, conditioned on the prosody and speech tokens of previous syllables as follows:
\begin{equation}
\boldsymbol{y}_i=p\left({\mathcal C}^{\rm p} \mid {\boldsymbol v}, \boldsymbol{X}, {q}_{1:i-1}, {{\boldsymbol S}}_{1:i-1}\right),
\end{equation}

\noindent where $i= \left\{1,...,I\right\}$. For the first syllable where \(i=1\), the preceding prosody and speech tokens are not utilized in its prediction process. The notation \({\boldsymbol S}_{i}\) represents the speech token indexes of the frames in the \(i\)-th syllable, defined as \({\boldsymbol S}_{i} = \{{s}_{i,1}, {s}_{i,2}, \ldots, {s}_{i,T_i}\}\), with \(T_i\) denoting the number of frames in the \(i\)-th syllable. Particularly, a prosody query embedding (\texttt{PQ}) is used as input to trigger the prediction of a prosody token. The prosody token index of the $i$-th syllable $q_i$ is derived from $\boldsymbol{y}_i$.

Subsequently, the probability distribution over the speech token set of the $t$-th frame in the $i$-th syllable is predicted conditioned on the predicted prosody and speech tokens as follows.
\begin{equation}
\boldsymbol{z}_{i,t} = p\left(\mathcal{C}^{\rm s} \mid {\boldsymbol v}, \boldsymbol{X}, q_{1:i-1},{\boldsymbol{S}}_{1:i-1}, q_{i}, s_{i,1:t-1}\right),
\end{equation}
where $t = 1,...,T_{i}+1$, and ${\mathcal{C}}^{\rm s}$ is the speech token set, with the superscript $^{\rm s}$ short for speech. Particularly, within the speech tokens, end-of-syllable (\texttt{EOSL}) and end-of-sequence (\texttt{EOS}) tokens are used to indicate the end of a syllable and the entire utterance, respectively. Thus, $s_{i,T_i+1}$ represents the \texttt{EOSL} token for $i<I$, and $s_{I,T_I+1}$ represents the \texttt{EOS} token. For the first frame in the \(i\)-th syllable where \(t=1\), the prior speech tokens within the syllable are not applied. Finally, the speech token index $s_{i,t}$ is derived from ${\boldsymbol{z}}_{i,t}$.

\vspace{-0.2em}
\subsection{Prosody token}
\vspace{-0.2em}
\label{sec:token}
Given a speech utterance, the energy and pitch are extracted for each syllable. Four prosody-related features are computed for the $i$-th syllable: duration $d_i$, mean energy $e_i$, mean pitch $h_i$, and pitch range $r_i$. Specifically, $e_i$ and $h_i$ are computed as the mean of the corresponding features within the syllable, and $r_i$ is calculated as the difference between the maximum and minimum pitch values within the syllable. The syllable-level prosody vector is composed by combining these values as $ \boldsymbol{g}_i = \left[d_i, e_i, h_i, r_i\right] $. It is quantized using k-means clustering, with the prosody token for the $i$-th syllable $q_i$ obtained as its nearest cluster as follows:
\begin{equation}
q_i = \arg\min_{j} \lVert \boldsymbol{g}_i - \boldsymbol{\mu}_j \rVert^2,
\label{eq: prosody token}
\end{equation}
where $\boldsymbol{\mu}_j$ denotes the centroid of the $j$-th cluster.

\vspace{-0.2em}
\subsection{Loss function}
\vspace{-0.2em}
The loss function is defined by minimizing the cross-entropy losses computed on the prosody and speech tokens. Assume the ground-truth vectors for the prosody token of the $i$-th syllable to be $\hat{\boldsymbol{y}}_i$ and the speech token of the $t$-th frame in this syllable to be $\hat{\boldsymbol{z}}_{i,t}$. The loss function is formulated as follows:
\vspace{-0.6em}
\begin{equation}
\begin{split}
\mathcal{L} & = -\alpha\frac{1}{I}\sum_{i=1}^{I}\hat{\boldsymbol{y}}_i\log \boldsymbol{y}_i \\
& \quad -\left(1-\alpha\right)\frac{1}{\sum_{i=1}^{I} (T_i+1)}\sum_{i=1}^{I}\sum_{t=1}^{T_i+1}\hat{\boldsymbol{z}}_{i,t}\log \boldsymbol{z}_{i,t},
\end{split}
\end{equation}
\vspace{-0.6em}
where $0\le \alpha\le 1$ is the weight variable.

\vspace{-0.2em}
\subsection{Inference}
\vspace{-0.2em}

Fig. \ref{fig: proposed}(\subref{fig:sequence}) depicts the input and generated token sequence in the inference process. Given the reference speech and its text transcription, reference prosody tokens are derived according to Eq. (\ref{eq: prosody token}). Meanwhile, speech tokens for the reference speech are obtained. These tokens, combined with the speaker embedding \(\boldsymbol{v}\) extracted from the reference speech and the syllable-level text embedding of the reference and target texts, are fed into the LLM, generating the prosody and speech tokens alternately. Prosody token and speech token indices, \( q_i \) and \( s_{i,t} \), are sampled from the predicted distributions, \( \boldsymbol{y}_i \) and \( \boldsymbol{z}_{i,t} \), using top-$p$ and top-$k$ sampling strategies. Finally, the speech tokens are input into the flow matching module to synthesize the speech waveform.

\vspace{-0.6em}
\section{Experiments}
\vspace{-0.2em}

\vspace{-0.2em}
\subsection{Datasets}
\vspace{-0.2em}

Experiments were conducted on Mandarin Chinese, in which a character corresponds to a single syllable. The training data was derived from WenetSpeech \cite{zhang2022wenetspeech}, which contains 10k hours of speech, and the Mandarin subset of Emilia \cite{he2024emilia}, which contains 50k hours of speech. Syllable boundaries were determined using the Montreal forced aligner (MFA) \cite{mcauliffe2017montreal}. By removing MFA failures and language-misclassified speech, approximately 50k hours of data were used for training.

Three datasets were used for evaluation, including: 1) ESD\cite{zhou2021seen}: 350 utterances from 10 speakers expressing 5 emotions (neutral, happy, angry, sad, surprised), rich in emotional prosody; 2) internal: 230 unlabeled utterances with diverse speaking styles, provided by iFLYTEK; 3) AISHELL-3\cite{shi2020aishell}: the test set in AISHELL-3, containing prosodically neutral recordings from 214 speakers. These test sets provided stylistic diversity (ESD and internal) and prosodic neutrality (AISHELL-3) for a comprehensive evaluation.

During inference in the ESD dataset, a reference speech and its text transcription were assigned a target text, which was recorded by the same speaker in the same emotion. In the AISHELL-3 dataset, the target text adopted the transcription of a different utterance by the same speaker as the reference speech. In the internal test set, the target text was randomly selected from a different utterance from the reference speech.

\vspace{-0.2em}
\subsection{Compared methods}
\vspace{-0.2em}

Our experiments employed the CosyVoice architecture using its open-source implementation\footnote{\label{cosyvoice_github}\url{https://github.com/FunAudioLLM/CosyVoice}} along with the same speech tokenizer and configurations. Speaker embedding was extracted by a pre-trained speaker encoder CAM++\cite{cam++}\footnote{\label{spk_ebd}\url{https://www.modelscope.cn/models/iic/speech_campplus_sv_zh-cn_16k-common}}. 
To evaluate the proposed prosody prediction method's efficacy, we compared three models trained on our 50k hours dataset:

\noindent \romannumeral 1. \emph{CosyVoice(50k)}: A CosyVoice LLM was trained from scratch using the 50k dataset.

\noindent \romannumeral 2. \emph{CoT}: Based on the CosyVoice LLM framework, given the target text and a reference speech combined with its text, the prosody tokens for the entire target speech were predicted at the syllable level.

\noindent \romannumeral 3. \emph{Proposed}: The CosyVoice LLM trained with the proposed dynamic prosody prediction. The weight variable $\alpha$ was set to 0.5.


The k-means clusters were trained on the WenetSpeech dataset to extract prosody tokens, using a cluster size of 512. The clusters were used in both the CoT and proposed methods. The LLM employed a decoder-only Transformer architecture, which contained 14 transformer layers with 16 attention heads, 1024 embedding dimensions, and 4096 feed-forward dimensions. All the models were trained for 800,000 steps on eight MLU 580 GPUs, with a learning rate of $l_r=10^{-4}$ and a warmup step setting of 10,000. During inference, we set top-$p$ sampling to $p=0.8$ with top-$k$ constraints: $k=25$ for speech tokens and $k=15$ for prosody tokens.

\begin{table}[t]
\setlength{\tabcolsep}{10pt}
\centering
\caption{MOS evaluation with 95\% confidence intervals among recording, speech generated by CosyVoice(50k), CoT and Proposed methods.}
\vspace{-0.8em}
\label{tab:mos}
\scalebox{0.8}{
\begin{tabular}{lcccc}
\toprule
\multirow{2}{*}{Method} & \multicolumn{3}{c}{MOS} \\
\cmidrule(lr){2-4}
& ESD & Internal & AISHELL-3 \\
\midrule

Recording
& $4.21\pm 0.09$ & $4.16\pm 0.07$ & $4.18\pm 0.08$ \\

\midrule

CosyVoice(50k)
& $4.01\pm 0.07$ & $3.97\pm 0.07$ & $4.06\pm 0.06$ \\

CoT
& $4.00\pm 0.09$ & $3.98\pm 0.08$ & $4.03\pm 0.10$ \\

Proposed
& $4.07 \pm 0.06$ & $3.99\pm 0.07$ & $4.06 \pm 0.07$ \\

\bottomrule
\end{tabular}}
\end{table}

\begin{table}[t]
\centering
\small
\caption{Average preference scores between CosyVoice(50k) and CoT methods (Method A) and the proposed model (Method B) on three evaluation datasets, where N/P means ``no preference''.}
\vspace{-0.8em}
\label{tab:preference}
\scalebox{0.8}{
\begin{tabular}{lcccc}
\toprule
Dataset & Method A & Prefer A(\%) & N/P(\%) & Prefer B(\%) \\
\midrule

\multirow{2}{*}{ESD}
& CosyVoice(50k) & 28.8 & 19.7 & \textbf{51.5} \\
& CoT & 28.8 & 21.4 & \textbf{50.9} \\
\midrule

\multirow{2}{*}{Internal}
& CosyVoice(50k) & 33.2 & 18.6 & \textbf{48.2} \\
& CoT & 30.9 & 21.4 & \textbf{47.7} \\
\midrule

\multirow{2}{*}{AISHELL-3}
& CosyVoice(50k) & 20.4 & 45.9 & \textbf{33.6} \\
& CoT & 25.9 & 40.5 & \textbf{33.6} \\
\bottomrule
\vspace{-2em}
\end{tabular}}
\end{table}

\begin{table*}[!t]
\setlength{\tabcolsep}{8pt}
\centering
\vspace{-2em}
\caption{Objective evaluation results of the compared models across the three datasets.}
\vspace{-0.8em}
\label{tab:objective2}
\scalebox{0.9}{
\begin{tabular}{lcccccccc}
\toprule
\multirow{2}{*}{{Dataset}} & \multirow{2}{*}{{Model}} & \multirow{2}{*}{{CER}} &  \multicolumn{2}{c}{{Emotion}} & \multicolumn{2}{c}{{Pitch}} & \multicolumn{2}{c}{{Energy}} \\
\cmidrule(lr){4-5} \cmidrule(lr){6-7} \cmidrule(lr){8-9}
& & &  {SIM} $\uparrow$ & {ACC(\%)} $\uparrow$ & {Corr(\%)} $\uparrow$ & {RMSE} $\downarrow$ & {Corr(\%)} $\uparrow$ & {RMSE} $\downarrow$\\
\midrule

\multirow{3}{*}{ESD}
& CosyVoice(50k) & 6.38 & 0.875 & 84.32 & 79.52 & 83.61 & 94.08 & 6.42 \\
& CoT & 6.14 & 0.876 & 84.52 & 79.31 & 82.82 & 94.03 & 6.39 \\
& Proposed & \textbf{5.66} & \textbf{0.884} & \textbf{86.56} & \textbf{80.32} & \textbf{80.81} & \textbf{94.91} & \textbf{5.93} \\
\midrule

\multirow{3}{*}{Internal}
& CosyVoice(50k)  & 13.69 & 0.802 & \textbf{52.31} &  &  &  &  \\
& CoT  & 13.6 & 0.799 & 50.23 & \multicolumn{4}{c}{------} \\
& Proposed  & \textbf{10.44} & \textbf{0.821} & \underline{51.63} &  &  &  &  \\
\midrule
\multirow{3}{*}{AISHELL-3}
& CosyVoice(50k) & 11.59 &  &  & 80.41 & 69.92 & 90.59 & 6.66 \\
& CoT & 11.61 & \multicolumn{2}{c}{------} & 80.61 & 69.90 & 90.51 & 6.52 \\
& Proposed &  \textbf{10.19} &  &  & \textbf{82.58} & \textbf{66.08} & \textbf{92.66} & \textbf{5.91} \\

\bottomrule

\vspace{-2em}
\end{tabular}}
\end{table*}

\vspace{-0.4em}
\subsection{Subjective evaluation}
\vspace{-0.2em}

The mean opinion score (MOS) and preference tests were conducted by eleven native speakers to evaluate naturalness and speaker similarity, respectively. A test set of 20 sentences was randomly selected from each test set. In the preference tests, the synthesized utterances were concatenated with the reference recording, respectively. Judges were asked to determine which concatenated utterance exhibited higher internal consistency throughout the entire sequence. The evaluation measured the speaker consistency and similarity between the reference and synthesized utterances in two dimensions: voice, speaking style (specifically emotion in the ESD test set). Higher preference scores indicated greater speaker similarity between the reference and generated speech.

The MOS test results are presented in Table \ref{tab:mos}. The proposed method achieved comparable results on the ESD, internal and AISHELL-3 datasets, indicating that the proposed method did not degrade the naturalness of the synthesized speech. Results of preference tests obtained at $p < 0.01$ are presented in Table \ref{tab:preference}. The results indicate that utterances synthesized using our proposed method received higher preference scores compared to those synthesized with the other two methods, demonstrating its efficacy in enhancing speaker similarity.

\vspace{-0.2em}
\subsection{Objective evaluation}
\vspace{-0.2em}

Objective evaluations were conducted to further analyze the intelligibility and the prosody modeling capability. The intelligibility was measured by character error rate (CER) obtained by Whisper\cite{radford2023robust} ASR model. To further analyze the capability of the proposed method in prosody prediction, emotion evaluation and prosodic feature assessment were performed.
The emotion evaluation was examined by the cosine similarity (SIM) between the emotion embeddings derived from the synthesized and reference speech, and the accuracy (ACC) of emotion recognition on the synthesized speech. The speech emotion recognition model emotion2vec+ \cite{ma2023emotion2vec}\footnote{\label{ser}\url{https://github.com/ddlBoJack/emotion2vec}} was used for both emotion embedding extraction and emotion recognition. In the prosodic feature assessment, the correlation (Corr) and root mean square error (RMSE) were computed on the pitch and energy values between the generated speech and recording. In each test set, 1000 utterances were generated for evaluation. The results are shown in Table \ref{tab:objective2}. Notably, the internal dataset was not assessed for prosodic features due to the lack of speaker labels. 
The AISHELL-3 dataset was excluded from emotional evaluation due to its prosodically neutral characteristics.


Table \ref{tab:objective2} shows the objective results. The proposed method achieves lower CERs compared to CosyVoice(50k) and CoT across the three test sets, indicating that the proposed method did not degrade the linguistic content of the synthesized speech.
In the emotion evaluation, the proposed method outperformed the other two methods in both emotion similarity and recognition accuracy on the ESD test set.
On the internal test set, it attained the highest emotion similarity and the second-highest recognition accuracy. These results indicate the efficacy of the proposed method to learn and transfer emotion from the reference speech to the synthesized speech. Moreover, in the prosodic feature evaluations, the proposed method exhibited the highest correlations and the lowest RMSE for both pitch and energy compared to the other two methods, further demonstrating its superior capability in prosody learning. The superiority in prosody learning of the proposed method endows it with better capability to learn the speaking style of the target speaker, thereby enhancing the speaker similarity of the generated speech.


\begin{table}[t]
\setlength{\tabcolsep}{3pt}
\centering
\caption{Average preference scores between Vevo1.5, F5-TTS and CosyVoice models (Model A) and the proposed model (Model B) on three evaluation datasets, where N/P means ``no preference''. The p-value of each test is included.}
\vspace{-0.8em}
\label{tab:preference2}
\scalebox{0.9}{
\begin{tabular}{lccccc}
\toprule
{Dataset} & {Model A} & {Prefer A(\%)} & {N/P(\%)} & {Prefer B(\%)} & $p$ \\
\midrule

\multirow{3}{*}{ESD}
& Vevo1.5 & 27.8 & 19.4 & \textbf{52.8} & $<0.01$ \\
& F5-TTS & 22.7 & 26.4 & \textbf{50.9} & $<0.01$ \\
& CosyVoice & 32.7 & 22.4 & \textbf{44.8} & $<0.01$ \\
\midrule

\multirow{3}{*}{Internal}
& Vevo1.5 & 36.4 & 16.6 & \textbf{47.0} & $<0.01$  \\
& F5-TTS & 24.1 & 15.5 & \textbf{60.4} & $<0.01$ \\
& CosyVoice & 35.4 & 26.4 & 38.2 & $0.10$ \\
\midrule

\multirow{3}{*}{AISHELL-3}
& Vevo1.5 & 23.2 & 41.2 & \textbf{35.6} & $<0.01$ \\
& F5-TTS & 21.4 & 50.0 & \textbf{28.6} & $<0.01$ \\
& CosyVoice & 29.1 & 43.2 & 27.7 & $0.65$ \\
\bottomrule
\vspace{-2em}
\end{tabular}}
\end{table}

\vspace{-0.4em}
\subsection{Comparison with open-source models}
\vspace{-0.2em}

Finally, we compared our proposed model (trained on a 50k-hour dataset) with three representative open-sourced TTS models to evaluate its competitive performance. These models include: 1) the implicit prosody modeling based CosyVoice and F5-TTS \cite{chen2024f5}\footnote{\label{f5_tts}\url{https://github.com/SWivid/F5-TTS}}; and 2) the explicit prosody modeling based Vevo1.5 \cite{vevo}\footnote{\label{vevo1.5}\url{https://github.com/open-mmlab/Amphion}}, which employs CoT prompting for prosody modeling.
Specifically, the CosyVoice model was trained on a multilingual corpus totaling 170k hours, including 130k hours of Chinese data. Both the Vevo1.5 and F5-TTS models were trained with 100k hours of the Emilia dataset, comprising 50k hours of Chinese and 50k hours of English speech.

Preference tests were conducted between each of the large models with the proposed model on the three test sets. The results are shown in Table \ref{tab:preference2}. The proposed method outperformed CosyVoice on prosodically rich datasets (the ESD and internal test sets), suggesting that the proposed dynamic prosody prediction method was able to enhance the CosyVoice architecture's performance in prosody modeling with a reduced training dataset. Besides, the proposed model exhibited superior performance compared to the Vevo1.5 and F5-TTS models across the three test sets. This confirms the advantages of dynamic prediction over both implicit and CoT-based modeling paradigms. These results further imply the potential of the proposed mechanism to serve as a robust compensator for the prosody learning deficit inherent in small-scale training, effectively narrowing the performance gap between limited and large-scale data regimes.

\vspace{-0.2em}
\section{Conclusions}
\vspace{-0.2em}

This paper aims to enhance speaker similarity in personalized TTS through dynamic prediction of stylistic prosody specific to target speech expressions.
Building upon the LLM-based TTS framework, we propose a dynamic prosody prediction method that predicts the prosody of the current syllable by considering the previously predicted speech. 
Evaluations on the ESD dataset, an internal dataset, and AISHELL-3 dataset demonstrated the effectiveness of the proposed dynamic prosody prediction method in enhancing prosody learning capabilities and thereby improving speaker similarity. The proposed prosody prediction strategy exhibited the potential to reduce the prosody learning gap between small- and large-scale training data, which highlights the need for future investigation.


\clearpage

\bibliographystyle{IEEEtran}
\bibliography{refs}

\end{document}